\documentclass[a4paper,11pt]{article}
\pdfoutput=1 

\usepackage{jcappub} 

\usepackage[T1]{fontenc} 

\newcommand{\prl}{Phys. Rev. Lett.}
\newcommand{\prd}{Phys. Rev. D}
\newcommand{\apjl}{Astrophys. J. Lett.}
\newcommand{\apj}{Astrophys. J.}
\newcommand{\aap}{Astron. Astrophys.}
\newcommand{\nat}{Nature}
\newcommand{\jcap}{J. Cosmol. Astropart. Phys.}

\title{\boldmath Bounding the Photon Mass with the Dedispersed Pulses of the Crab Pulsar and FRB 180916B}

\author[a,b]{Chen-Ming Chang,}
\author[a,b,1]{Jun-Jie Wei,\note{Corresponding author.}}
\author[a,1]{Song-bo Zhang}
\author[a,b,1]{and Xue-Feng Wu}

\affiliation[a]{Purple Mountain Observatory, Chinese Academy of Sciences,\\Nanjing 210023, China}
\affiliation[b]{School of Astronomy and Space Sciences, University of Science and Technology of China,\\Hefei 230026, China}

\emailAdd{jjwei@pmo.ac.cn}
\emailAdd{sbzhang@pmo.ac.cn}
\emailAdd{xfwu@pmo.ac.cn}

\abstract{Tight limits on the photon mass have been set through analyzing the arrival time differences of photons with different
frequencies originating from the same astrophysical source. However, all these constraints have relied on using
the first-order Taylor expansion of the dispersion due to a nonzero photon mass. In this work, we present an analysis
of the nonzero photon mass dispersion with the second-order derivative of Taylor series. If the arrival time delay corrected
for all known effects (including the first-order delay time due to the plasma and photon mass effects) is assumed to be
dominated by the second-order term of the nonzero photon mass dispersion, a conservative upper limit on the photon mass can be estimated. Here we show that the dedispersed pulses with the second-order time delays from the Crab pulsar and 
the fast radio burst FRB 180916B pose strict limits on the photon mass, i.e.,
$m_{\gamma,2} \leq5.7\times10^{-46}\;{\rm kg}\simeq3.2\times10^{-10}\; {\rm eV}/c^{2}$
and $m_{\gamma,2} \leq6.0\times10^{-47}\;{\rm kg}\simeq3.4\times10^{-11}\; {\rm eV}/c^{2}$, respectively.
This is the first time to study the possible second-order photon mass effect.}


\arxivnumber{2207.00950}

\begin{document}
\maketitle
\flushbottom

\section{INTRODUCTION}
\label{sec:intro}

A basic postulate of Maxwell's electromagnetism as well as Einstein's theory of special relativity is the constant speed $c$,
in vacuum, of all electromagnetic radiation, which implies that the photon should be massless. Determining the rest mass of
the photon has therefore been one of the most enduring efforts on testing the validity of this postulate. However, it is
unfeasible to prove experimentally that the photon rest mass is exactly zero. According to the uncertainty principle,
the ultimate upper limit on the photon rest mass would be $m_\gamma \leq \hbar / \Delta t c^2 \approx {10}^{-69}$ kg, using
the age of the universe of about $10^{10}$ years \citep{1971RvMP...43..277G,2005RPPh...68...77T}. The optimal experimental
strategy is therefore to set ever tighter upper bounds on $m_\gamma$ and push the results more closely towards the ultimate
bounds of measurement uncertainty.

From a theoretical perspective, a nonzero photon rest mass can be accommodated in a unique way by changing the inhomogeneous
Maxwell's equations to the Proca equations. Using them, it is possible to consider some far-reaching implications associated
with massive photons, such as variations of the speed of light with frequency, deviations in the behavior of static
electromagnetic fields, the existence of longitudinal electromagnetic waves, and so on. All of these effects have been
employed to set upper limits on the photon rest mass via various terrestrial and extra-terrestrial approaches
\citep{1971RvMP...43..277G,1973PhRvD...8.2349L,1997sref.book.....Z,2005RPPh...68...77T,2006AcPPB..37..565O,2010RvMP...82..939G,
2011EPJD...61..531S}. Over the past several decades, the common approaches for determining the photon mass include
measurement of the frequency dependence in the speed of light \citep{1964Natur.202..377L,1969Natur.222..157W,
PhysRevLett.82.4964,Wu_2016,2016PhLB..757..548B,2017PhLB..768..326B,ZHANG201620,PhysRevD.95.123010,Wei_2017,
2018JCAP...07..045W,2020RAA....20..206W,2021FrPhy..1644300W,Xing_2019,2021PhLB..82036596W}, null tests of Coulomb's
inverse square law \citep{PhysRevLett.26.721}, tests of Amp\`ere's law \citep{PhysRevLett.68.3383}, torsion balance
\citep{PhysRevLett.80.1826,2003PhRvL..90h1801L}, Jupiter magnetic field \citep{PhysRevLett.35.1402}, magnetohydrodynamic
phenomena of the solar wind \citep{1997PPCF...39...73R,2007PPCF...49..429R,2016APh....82...49R}, cosmic magnetic fields
\citep{1959PThPS..11....1Y,1976UsFiN.119..551C,2007PhRvL..98a0402A},  gravitational deflection of massive photons
\citep{PhysRevD.8.2349,PhysRevD.69.107501}, suppermassive black hole spin \citep{PhysRevLett.109.131102}, pulsar spindown
\citep{Yang_2017}, and so on. Among these approaches, the most direct one is to measure a possible frequency dependence
in the velocity of light.

According to Einstein's special relativity, the energy of the photon with a nonzero rest mass $m_\gamma$ can be written as
\begin{equation}
E=h\nu=\sqrt{p^2c^2+m^2_\gamma c^4}\;.
\end{equation}
The relation between the massive photon group velocity $\upsilon$ and the frequency
$\nu$ then takes the form:
\begin{equation}\label{velocity}
\upsilon=\frac{\partial E}{\partial p}=c\sqrt{1-\frac{m^2_\gamma c^4}{E^2}}=c\sqrt{1-A\nu^{-2}}\;,
\end{equation}
where $A=m^2_\gamma c^4/h^2$. It is evidently clear from Equation (\ref{velocity}) that the lower frequency, the slower
the massive photon propagates in vacuum. The photon mass can therefore be constrained by comparing the arrival-time differences
of photons with different frequencies originating from the same source. Moreover, it is easy to understand that measurements
of shorter arrival-time differences between lower energy bands from sources at longer propagation distances are particularly
sensitive to the photon mass. Thanks to their fine time structures, radio emissions, and cosmological distances,
fast radio bursts (FRBs) provide the current best celestial laboratory for constraining the photon mass
\citep{Wu_2016,2016PhLB..757..548B,2017PhLB..768..326B,PhysRevD.95.123010,Xing_2019,2020RAA....20..206W,2021FrPhy..1644300W,
2021PhLB..82036596W}. The first attempts to place upper limits on the photon mass using FRBs were presented in refs.~\cite{Wu_2016,2016PhLB..757..548B}. Adopting the controversial redshift $z=0.492$ for FRB 150418, ref.~\cite{Wu_2016} set a strict
upper limit of $m_\gamma \leq 5.2 \times {10}^{-50}$ kg (see also ref.~\cite{2016PhLB..757..548B}). Subsequently,
ref.~\cite{2017PhLB..768..326B} used the reliable redshift measurement of FRB 121102 to derive a robust limit of
$m_\gamma \leq 3.9 \times {10}^{-50}$ kg. Ref.~\cite{PhysRevD.95.123010} developed a Bayesian framework to constrain the photon mass
with a catalog of 21 FRBs, yielding $m_\gamma \leq 8.7 \times {10}^{-51}$ kg at $68\%$ confidence level. By analyzing
the time-frequency structure of subbursts in FRB 121102, ref.~\cite{Xing_2019} obtained a tighter limit on the photon mass of
$m_\gamma \leq 5.1 \times {10}^{-51}$ kg. Since the plasma and photon mass contributions to the dispersion measure (DM) have
different redshift dependences, ref.~\cite{2020RAA....20..206W} proved that they can be distinguished by measurements of nine FRB
redshifts, enabling the sensitivity to $m_\gamma$ to be improved (i.e., $m_\gamma \leq 7.1 \times {10}^{-51}$ kg at $68\%$
confidence level). Ref.~\cite{2021PhLB..82036596W} used a catalog of 129 FRBs in a Bayesian framework to derive a combined limit
of $m_\gamma \leq 3.1 \times {10}^{-51}$ kg at $68\%$ confidence level.

Although photon mass limits obtained through the dispersion method have reached high precision, all current investigations
considered the first-order Taylor expansion of the dispersion relation only
(i.e., $\Delta t_{m_{\gamma},1}\propto {m^{2}_\gamma}/{\nu^2}$).
If we keep the second-order derivative of Taylor series, the frequency-dependent dispersion due to a nonzero photon
mass would be $\Delta t_{m_{\gamma}}=\Delta t_{m_{\gamma},1}+\Delta t_{m_{\gamma},2}\propto a_{1}{m^{2}_\gamma}/{\nu^2}+a_{2}{m^{4}_\gamma}/{\nu^4}$.
From observations, all radio emission signatures show an indisputable $\nu^{-2}$-dependent time delay,
$\Delta t_{\rm obs,1}\propto {\rm DM}\cdot\nu^{-2}$,
which is in good agreement with the first-order Taylor expansion of the dispersions expected from both the plasma
($\Delta t_{\rm DM,1}$) and nonzero photon mass ($\Delta t_{m_{\gamma},1}$, if it exists) effects. That is,
$\Delta t_{\rm obs,1}=\Delta t_{\rm DM,1}+\Delta t_{m_{\gamma},1}$. Assuming that the dedispersed time delay
($\Delta t_{\rm obs}-\Delta t_{\rm obs,1}$) is attributed solely to the second-order term arising from the nonzero
photon mass effect ($\Delta t_{m_{\gamma},2}$), we can therefore obtain a new upper limit on the photon mass
$m_{\gamma,2}$.

With the dedispersed time delays from radio observations in hand, it is interesting to investigate what level of
$m_{\gamma,2}$ limits can be obtained by taking into account the second-order Taylor expansion of the dispersion due to
a nonzero photon mass. In this work, we make use of the dedispersed pulses from the Crab pulsar \citep{Hankins_2007} and
FRB 180916B \citep{2021ApJ...911L...3P}, for the first time, to study the possible second-order photon mass effect.
Throughout this paper a flat $\Lambda$CDM cosmological model with $H_{0}=67.36$ km $\rm s^{-1}$ $\rm Mpc^{-1}$, 
$\Omega_m=0.315$, and $\Omega_{\Lambda}=0.685$ is adopted \citep{2020A&A...641A...6P}.

\section{Theoretical Framework}
\subsection{Dispersion from a Nonzero Photon Mass}

With Equation (\ref{velocity}), it is straightforward to show that two massive photons emitted simultaneously from
a same source would reach us at different times if they have different frequencies. For a cosmic source at redshift $z$,
the rest-frame time delay ($\Delta{t_{m_{\gamma},z}}$) between two massive photons with different rest-frame frequencies
($\nu_{l,z}<\nu_{h,z}$) can be obtained by using second-order approximation of Taylor expansion:
\begin{equation}\label{eq:tmrz}
\begin{aligned}
  \Delta{t_{m_{\gamma},z}}&=\int \frac{{\rm d}l}{c}\left[\left(1-A\nu_{l,z}^{-2}\right)^{-1/2}-\left(1-A\nu_{h,z}^{-2}\right)^{-1/2}\right]\\
   &\simeq\int \frac{{\rm d}l}{c} \left[\frac{1}{2}A\left(\nu_{l,z}^{-2}-\nu_{h,z}^{-2}\right)+\frac{3}{8}A^{2}\left(\nu_{l,z}^{-4}-\nu_{h,z}^{-4}\right)\right]\;.
\end{aligned}
\end{equation}
In the observer frame, the observed time delay is $\Delta{t_{m_{\gamma}}}=\Delta{t_{m_{\gamma},z}}\times(1+z)$ and
the observed frequency is $\nu=\nu_{z}/(1+z)$. Noticing
\begin{equation}
{\rm d}l=\frac{1}{1+z}\frac{c}{H_0}\frac{{\rm d}z}{\sqrt{\Omega_{m}(1+z)^3+\Omega_{\Lambda}}}
\end{equation}
for a flat $\Lambda$CDM cosmological model, so Equation~(\ref{eq:tmrz}) can be modified as
\begin{equation}
\begin{aligned}
\Delta t_{m_\gamma}&=\frac{A}{2H_0}\left(\nu_l^{-2}-\nu_h^{-2}\right)H_1(z)+\frac{3A^2}{8H_0}\left(\nu_l^{-4}-\nu_h^{-4}\right)H_2(z)\\
&=\Delta t_{m_{\gamma},1}+\Delta t_{m_{\gamma},2}\;,
\end{aligned}
\end{equation}
where
\begin{equation}
H_1(z)=\int_0^z\frac{(1+z')^{-2}dz'}{\sqrt{\Omega_m(1+z')^3+\Omega_\Lambda}}
\end{equation}
and
\begin{equation}
H_2(z)=\int_0^z\frac{(1+z')^{-4}dz'}{\sqrt{\Omega_m(1+z')^3+\Omega_\Lambda}}\;.
\end{equation}

For a nearby source at distance $d$, we do not need to take into account the cosmological expansion. The observed time delay
induced by the nonzero photon mass effect can be simplified as
\begin{equation}
\begin{aligned}
\Delta t_{m_\gamma}&=\frac{d}{\upsilon_l}-\frac{d}{\upsilon_h}\\
&\approx \frac{dA}{2c}\left(\nu^{-2}_l-\nu^{-2}_h\right)+\frac{3dA^2}{8c}\left(\nu^{-4}_l-\nu^{-4}_h\right)\\
&=\Delta t_{m_{\gamma},1}+\Delta t_{m_{\gamma},2}\;,
\end{aligned}
\end{equation}
where $\Delta t_{m_{\gamma},1}=\frac{dA}{2c}\left(\nu^{-2}_l-\nu^{-2}_h\right)$ and
$\Delta t_{m_{\gamma},2}=\frac{3dA^2}{8c}\left(\nu^{-4}_l-\nu^{-4}_h\right)$ correspond to
the first- and second-order terms, respectively.

\subsection{Dispersion from the Plasma Effect}

Due to the dispersive nature of plasma, the group velocity of electromagnetic waves propagating
through the ionized median would have a frequency dependence, i.e.,
\begin{equation}
\upsilon_p=c\sqrt{1-\frac{{\nu_p}^2}{\nu^2}}\;,
\end{equation}
where the plasma frequency $\nu_{p}=(n_{e}e^{2}/4\pi^{2} m_{e}\epsilon_{0})^{1/2}$ with
$n_{e}$ the electron number density in the plasma, $m_{e}$ and $e$ the mass and charge of
an electron, respectively, and $\epsilon_{0}$ the permittivity. The arrival of a radio signal
with frequency $\nu$ travelling across the plasma along the line of sight would be delayed by
\citep{1979rpa..book.....R}
\begin{equation}
\tau=\int\frac{{\rm d}l}{c}\left[\left(1-\frac{{\nu_p}^2}{\nu^2}\right)^{-1/2}-1\right]
\end{equation}
with respect to the arrival time had the signal travelled through vacuum.
The delay time between two wave packets with different frequencies, which caused by the plasma effect,
can then be obtained by using second-order approximation of Taylor expansion:
\begin{equation}\label{eq:tDM}
\begin{aligned}
  \Delta{t_{\rm DM}}&\simeq\int \frac{{\rm d}l}{c} \left[\frac{1}{2}\nu^{2}_{p}\left(\nu_l^{-2}-\nu_h^{-2}\right)+\frac{3}{8}\nu^{4}_{p}\left(\nu_l^{-4}-\nu_h^{-4}\right)\right]\\
  &=\Delta{t_{{\rm DM},1}}+\Delta{t_{{\rm DM},2}}\;,
\end{aligned}
\end{equation}
where $\Delta{t_{{\rm DM},1}}$ and $\Delta{t_{{\rm DM},2}}$ correspond to the first- and second-order
terms, respectively. The first-order term $\Delta{t_{{\rm DM},1}}$ can be further reduced to
\begin{equation}
  \Delta{t_{{\rm DM},1}}=\frac{e^{2}}{8\pi^{2} m_{e}\epsilon_{0}c}\left(\nu_l^{-2}-\nu_h^{-2}\right){\rm DM}\;,
\end{equation}
where ${\rm DM}=\int n_{e}{\rm d}l$ is the dispersion measure, which is given in the absence of a photon mass
by the integrated electron column density along the propagation path. In a cosmological setting,
${\rm DM}=\int n_{e,z}(1+z)^{-1}{\rm d}l$, where $n_{e,z}$ is the electron number density in the rest-frame
and $z$ is the redshift \citep{Deng_2014}.

\subsection{The Residual/dedispersed Time Delay}

In our analysis, we suppose that the total time delay of two photons with different frequencies is attributed to
the following components:
\begin{equation}
\label{obs_delay}
\Delta t_{\rm obs} = \Delta t_{\rm int} + \Delta{t_{{\rm DM},1}}+\Delta{t_{{\rm DM},2}} + \Delta t_{m_{\gamma},1}+\Delta t_{m_{\gamma},2}\;,
\end{equation}
where $\Delta t_{\rm int}$ is the intrinsic time delay that depends on the geometry and radiation processes of the source, and
both $\Delta t_{\rm DM}$ and $\Delta t_{m_\gamma}$ include the first- and second-order terms.
In practice, the de-dispersion could remove all the time delay of $\nu^{-2}$ behavior, i.e., the first-order terms in Equation~(\ref{obs_delay}).
Thus, after dedispersed, the residual delay time is
\begin{equation}
\begin{aligned}
\Delta t_{\rm res}&=\Delta t_{\rm obs}-\Delta{t_{{\rm DM},1}}-\Delta t_{m_{\gamma},1}\\
&= \Delta t_{\rm int} +\Delta{t_{{\rm DM},2}} +\Delta t_{m_{\gamma},2}\;,
\end{aligned}
\end{equation}
If we assume that $\Delta t_{\rm int}+\Delta{t_{{\rm DM},2}}\geq 0$, then
$\Delta t_{m_{\gamma},2}=\Delta t_{\rm res}-(\Delta t_{\rm int}+\Delta{t_{{\rm DM},2}})\leq\Delta t_{\rm res}$.
Thus, $\Delta t_{\rm res}$ is the upper limit of $\Delta t_{m_{\gamma},2}$, providing an upper limit on the photon mass.

For a nearby source at distance $d$, the photon mass can then be constrained by the second-order effect as
\begin{equation}
m_{\gamma,2} \leq hc^{-2}\left[\frac{8c\Delta t_{\rm res}}{3d\left(\nu_l^{-4}-\nu_h^{-4}\right)}\right]^{1/4}\;,
\end{equation}
which can be simplified as
\begin{equation}\label{mass}
m_{\gamma,2}\leq\left(0.93\times{10}^{-46}\;{\rm kg}\right)\left\{\frac{\frac{\Delta t_{\rm res}}{\rm ns}}{\left[\left({\frac{\nu_l}{\rm GHz}}\right)^{-4}-\left({\frac{\nu_h}{\rm GHz}}\right)^{-4}\right]\frac{d}{\rm kpc}}\right\}^{1/4}\;.
\end{equation}

For a cosmic source at redshift $z$, the photon mass derived from the second-order effect is given by
\begin{equation}
m_{\gamma,2} \leq hc^{-2}\left[\frac{8H_0\Delta t_{\rm res}}{3\left(\nu_l^{-4}-\nu_h^{-4}\right)H_2(z)}\right]^{1/4}\;,
\end{equation}
which can be further reduced to
\begin{equation}\label{zmass}
m_{\gamma,2} \leq \left(2.02\times{10}^{-48}\;{\rm kg}\right)\left\{\frac{\frac{\Delta t_{\rm res}}{\rm ns}}{\left[\left({\frac{\nu_l}{\rm GHz}}\right)^{-4}-\left({\frac{\nu_h}{\rm GHz}}\right)^{-4}\right]H_2(z)}\right\}^{1/4}\;.
\end{equation}

\section{Second-order Photon mass limits from the dedispersed pulses}\label{sec:results}
Here we show that the nanosecond-long giant pulse observed from the Crab pulsar and the burst from FRB 180916B 
detected at very low frequencies (below 150\,MHz) can provide strict limits on the second-order photon mass $m_{\gamma,2}$.
Even though the Crab pulsar giant pulse is in our Milky Way galaxy, the effect of the ultra short duration (e.g., 
$\Delta t_{\rm res}\leq0.4$ ns) overcompensates the deficit in distance, and results in a stricter limit on $m_{\gamma,2}$ than extragalactic ms-duration FRBs observed at the similar frequency range ($\sim$ GHz). Furthermore,
even though the burst durations ($\sim40$--160 ms observed at 150\,MHz) of FRB 180916B are relatively large, 
the effect of the very-low-frequency emission overcompensates the deficit in duration, and leads to a much more stringent limit on $m_{\gamma,2}$ than the Crab pulsar giant pulse.

\subsection{The Crab Pulsar}\label{sec:results_1}
Giant pulses are one of the most conspicuous phenomena of radio pulsars, with extremely high fluxes (exceeding MJy) and
very short durations. Their typical durations are a few microseconds, but occasional pulses shorter than one nanosecond
(the so-called ``nanoshots'') have also been detected \citep{Hankins_2007}. To date, they have been observed from
the Crab pulsar \citep{2004ApJ...612..375C,2005AdSpR..35.1166J,2010A&A...524A..60J,Hankins_2007}
and some other pulsars \citep{2001ApJ...557L..93R,2003ApJ...590L..95J,2004ApJ...616..439S,2005ApJ...625..951K}.
Ref.~\cite{Hankins_2007} reported a giant pulse from the Crab pulsar that showed an extremely intense nanoshot with a flux
exceeding 2 MJy and an unresolved dedispersed duration, $\Delta t_{\rm res}\leq0.4$ ns. This pulse was recorded at 9.25 GHz
center frequency over a 2.2 GHz bandwidth. The distance of the Crab pulsar is $d=2$ kpc. With the above information of
the 0.4-nanosecond giant pulse of the Crab pulsar, a severe limit on the photon mass from Equation~(\ref{mass}) is
\begin{equation}
m_{\gamma,2} \leq5.7\times10^{-46}\;{\rm kg}\simeq3.2\times10^{-10}\; {\rm eV}/c^{2}\;.
\end{equation}

Here we use the dedispersed duration as the upper limit of $\Delta t_{m_{\gamma},2}$ caused by the $\nu^{-4}$ term, 
which could be slightly larger considering the uncertainty of the DM fitting. Using the DM errorbar of 
${10}^{-5}\; {\rm pc~{cm}^{-3}}$ suggested in Ref.~\cite{Hankins_2007}, which equals to a time delay of 0.24 ns, 
a conservative dedispersed duration should be 0.64 ns. With this conservative dedispersed duration, the limit on 
the photon mass turns to be $m_{\gamma,2} \leq6.3\times10^{-46}\;{\rm kg}\simeq3.5\times10^{-10}\; {\rm eV}/c^{2}\;$. 
It is obvious that the DM errorbar has little effect on the photon mass limits.

\subsection{FRB 180916B}\label{sec:results_2}
FRB 180916B is a well-studied repeating FRB source, located at a redshift of $z=0.0337$ \citep{2020Natur.577..190M}.
Ref.~\cite{2021ApJ...911L...3P} reported on the lowest-frequency detection to date of 18 bursts from FRB 180916B, 
observed at 110--188 MHz with LOFAR. One of these bursts detected between 124.8 MHz and 185.7 MHz has a dedispersed 
duration of $\Delta t_{\rm res} < 84\; {\rm ms}$. 
Note that the radiation band of this burst (124.8--185.7 MHz) is fully covered by the observing band 
of LOFAR (110--188 MHz). It is therefore reasonable to assume that the largest time delay between 
different frequencies from FRB 180916B should be the dedispersed duration. 
From Equation~(\ref{zmass}), we can tighten the constraint on the photon mass to
\begin{equation}
m_{\gamma,2} \leq6.0\times10^{-47}\;{\rm kg}\simeq3.4\times10^{-11}\; {\rm eV}/c^{2}
\end{equation}
for the dedispersed burst of FRB 180916B. However, as the pulse spectral widths of pulsars are wide, 
it is nearly impossible for the Crab pulsar to find such a signal whose radiation band is fully covered by 
the observing bandwidth. Thus, there is a caveat that the dedispersed duration of the Crab pulsar giant pulse
may not be used as the upper limit of the time delay. 
Nevertheless, we can use the result of the Crab pulsar as a contrast with that obtained by FRB 180916B. The result obtained by the 0.4-nanosecond giant pulse of the Crab pulsar (8.15--10.35 GHz) is one order of magnitude worse than that of the tens of milliseconds burst of FRB 180916B (124.8--185.7 MHz). This suggests that the observed frequency plays a more important role than the duration in constraining the photon mass, 
a lower frequency emission leading to better constraints on $m_{\gamma,2}$.

Same as section~\ref{sec:results_1}, considering the DM errorbar of $0.006\; {\rm pc~{cm}^{-3}}$ 
offered in Ref.~\cite{2021ApJ...911L...3P}, which equals to a time delay of 1.1\;ms and makes the dedispersed
duration to be a conservative value of 85.1\;ms, the limit on the photon mass should be 
$m_{\gamma,2} \leq6.0\times10^{-47}\;{\rm kg}\simeq3.4\times10^{-11}\; {\rm eV}/c^{2}$. Again, the DM errorbar 
does not affect on our resulting constraints on $m_{\gamma,2}$.

\section{CONCLUSION AND DISCUSSIONS}
The frequency-dependent time delays of radio waves from astrophysical sources have been widely used to constrain the rest mass
of the photon. In this work, we make use of the dedispersed time delays between different energies from the Crab pulsar and
FRB 180916B, for the first time, to test the zero-mass hypothesis for the photon. This is a step forward in the investigation
of the photon mass, since all current studies considered the first-order Taylor expansion of the dispersion arising from
a nonzero photon mass only and the second-order effect $\Delta t_{m_{\gamma},2}$ has not yet been explored.

Assuming that the dedispersed time delay is mainly caused by the second-order term of the nonzero photon mass dispersion,
we place robust upper limits on the photon mass: $m_{\gamma,2} \leq5.7\times10^{-46}\;{\rm kg}$
(or equivalently $m_{\gamma,2} \leq3.2\times10^{-10}\; {\rm eV}/c^{2}$) for the 0.4-nanosecond giant pulse of the Crab pulsar and $m_{\gamma,2} \leq6.0\times10^{-47}\;{\rm kg}$ (or equivalently $m_{\gamma,2} \leq3.4\times10^{-11}\; {\rm eV}/c^{2}$) for the burst of the order of tens of milliseconds from FRB 180916B. It is notable that the ms-duration pulse of FRB 180916B detected at 116.9--188 MHz provides a much more stringent limit on $m_{\gamma,2}$, improving by at lest one order of magnitude from 
the result based on the 0.4-nanosecond giant pulse of the Crab Pulsar observed at 8.15--10.35 GHz.

In this work, we use the hypothesis that the de-dispersion could remove the time delay of $\nu^{-2}$ behaviour as a standard $\nu^{-2}$ form is assumed during the DM fitting of pulsar and FRB data. 
However, as the $\nu^{-4}$ term is much smaller than the first order term, they may be largely absorbed in the fitting. 
To analyse this effect,   
we consider a set of data consists of $n$ points $(\nu_n,t_n)$ with the relation between time and frequency:
\begin{equation}\label{-2+-4}
\frac{t}{\rm ms} = A_r\left(\frac{\nu}{\rm GHz}\right)^{-2}+B_r\left(\frac{\nu}{\rm GHz}\right)^{-4}\;,
\end{equation}
where the coefficients of the $\nu^{-2}$ and $\nu^{-4}$ terms both contain the effects of nonzero photon mass and plasma dispersion and their true values can hardly be achieved. However, if the effect of nonzero photon mass is neglected, the relation between $A_r$ and $B_r$ can be estimated through Equation~(\ref{eq:tDM}):
\begin{equation}\label{B/A}
\frac{B_r}{A_r}\simeq\nu_p^2\sim{10}^{-19}\;.
\end{equation}
When the standard $\nu^{-2}$ form
\begin{equation}\label{eq:standard}
\frac{t}{\rm ms} = A_f\left(\frac{\nu}{\rm GHz}\right)^{-2}\;
\end{equation}
is used to fit this set of data, the result will be 
\begin{equation}
A_f = \left[1+\frac{\sum_{i=1}^n \left(\frac{\nu_i}{\rm GHz}\right)^{-6}}{\sum_{i=1}^{n} \left(\frac{\nu_i}{\rm GHz}\right)^{-4}}\times{10}^{-19}\right]A_r
\end{equation}
through the least square curve fitting. It is obvious that the fitting result $A_f$ is larger than the real value $A_r$ because of the existence of the $\nu^{-4}$ term and it is necessary to evaluate this effect. Here we firstly chose the parameters of FRB 180916B to carry out our simulation: $A_r = 1447.40$, $B_r = 1447.40\times{10}^{-19}$ as suggested in Equation~(\ref{B/A}), $124.8\; {\rm MHz}$ as the lowest frequency, $185.7\; {\rm MHz}$ as the highest frequency, $0.781\; {\rm MHz}$ as the frequency resolution, $3.93\; {\rm ms}$ as the time resolution, and generated 77 data points ($n = 77$) following Equation~(\ref{-2+-4}). These data points were then fitted using Equation~(\ref{eq:standard}).
The best-fitting result is $A_f = 1447.40 \pm 0.01$. The time delay between the lowest and highest frequency channels caused by the fitting error is $\Delta t_{\rm err} \sim 0.45\; {\rm ms}$, which is much larger than that caused by the $\nu^{-4}$ term ($\Delta t_{\rm DM,2} = 4.6\times{10}^{-13}\; {\rm ms}$). For the Crab pulsar, we set $A_r = 235.60$, $B_r = 235.60\times{10}^{-19}$, $8.15\; {\rm GHz}$ as the lowest frequency, $10.35\; {\rm GHz}$ as the highest frequency, $0.4\; {\rm ns}$ as the time resolution, and generated 1024 points. The fitting result is $A_f = 235.60 \pm 2.4 \times {10}^{-17}$, and the related $\Delta t_{\rm err}$ and $\Delta t_{\rm DM,2}$ are $1.18 \times {10}^{-18} \;{\rm ms}$ and $3.16 \times {10}^{-21} \;{\rm ms}$, respectively.
Therefore, even when it is applied to very high quality observational parameters, the effect of the $\nu^{-4}$ term is negligible for the DM fitting and the fitting uncertainty is mainly caused by the time and frequency resolution of the dataset. It is also notable that the time delays caused by the fitting error and the $\nu^{-4}$ term are much smaller than the dedispersed duration that we use to constrain the photon mass. 

Our best limit is four orders of magnitude worse than previous limits
($m_{\gamma,1} \leq{10}^{-51}$ kg) obtained through the first-order Taylor expansion of the dispersion relation
\citep{2020RAA....20..206W,2021PhLB..82036596W}. While the dedispersed time delays of the Crab pulsar and FRB 180916B do not
currently have the best sensitivity to photon mass limits, there is nonetheless merit to the result. This is the first time to explore the possible second-order photon mass effect. More stringent constraints on $m_{\gamma,2}$ can be expected as our
analysis method is applied to larger numbers of nanoseconds-long astrophysical pulses observed at lower frequencies.



\acknowledgments
We are grateful to the anonymous referee for helpful comments and suggestions.
We also thank D. Xiao and J.-J. Geng for helpful discussions.
This work is partially supported by the National Key Research and Development Program of
China (2022SKA0130101), the National Natural Science Foundation of China
(grant Nos.~11725314 and 12041306), the Key Research Program of Frontier Sciences (grant No.
ZDBS-LY-7014) of Chinese Academy of Sciences, ACAMAR Postdoctoral Fellow, China Postdoctoral Science Foundation (Grant No. 2020M681758), and the Natural Science Foundation of Jiangsu Province (grant Nos. BK20210998 and BK20221562).










\providecommand{\href}[2]{#2}\begingroup\raggedright\endgroup


\begin{thebibliography}{10}

\bibitem{1971RvMP...43..277G}
A.S.~{Goldhaber} and M.M.~{Nieto}, \emph{{Terrestrial and Extraterrestrial
  Limits on The Photon Mass}},
  \href{https://doi.org/10.1103/RevModPhys.43.277}{\emph{Reviews of Modern
  Physics} {\bfseries 43} (1971) 277}.

\bibitem{2005RPPh...68...77T}
L.-C.~{Tu}, J.~{Luo} and G.T.~{Gillies}, \emph{{The mass of the photon}},
  \href{https://doi.org/10.1088/0034-4885/68/1/R02}{\emph{Reports on Progress
  in Physics} {\bfseries 68} (2005) 77}.

\bibitem{1973PhRvD...8.2349L}
D.D.~{Lowenthal}, \emph{{Limits on the Photon Mass}},
  \href{https://doi.org/10.1103/PhysRevD.8.2349}{\emph{\prd} {\bfseries 8}
  (1973) 2349}.

\bibitem{1997sref.book.....Z}
Y.Z.~{Zhang}, \emph{{Special Relativity and its Experimental Foundation}}
  (1997), \href{https://doi.org/10.1142/3180}{10.1142/3180}.

\bibitem{2006AcPPB..37..565O}
L.B.~{Okun}, \emph{{Photon: History, Mass, Charge}}, {\emph{Acta Physica
  Polonica B} {\bfseries 37} (2006) 565}
  [\href{https://arxiv.org/abs/hep-ph/0602036}{{\ttfamily hep-ph/0602036}}].

\bibitem{2010RvMP...82..939G}
A.S.~{Goldhaber} and M.M.~{Nieto}, \emph{{Photon and graviton mass limits}},
  \href{https://doi.org/10.1103/RevModPhys.82.939}{\emph{Reviews of Modern
  Physics} {\bfseries 82} (2010) 939}
  [\href{https://arxiv.org/abs/0809.1003}{{\ttfamily 0809.1003}}].

\bibitem{2011EPJD...61..531S}
G.~{Spavieri}, J.~{Quintero}, G.T.~{Gillies} and M.~{Rodr{\'\i}guez}, \emph{{A
  survey of existing and proposed classical and quantum approaches to the
  photon mass}},
  \href{https://doi.org/10.1140/epjd/e2011-10508-7}{\emph{European Physical
  Journal D} {\bfseries 61} (2011) 531}.

\bibitem{1964Natur.202..377L}
B.~{Lovell}, F.L.~{Whipple} and L.H.~{Solomon}, \emph{{Relative Velocity of
  Light and Radio Waves in Space}},
  \href{https://doi.org/10.1038/202377a0}{\emph{\nat} {\bfseries 202} (1964)
  377}.

\bibitem{1969Natur.222..157W}
B.~{Warner} and R.E.~{Nather}, \emph{{Wavelength Independence of the Velocity
  of Light in Space}}, \href{https://doi.org/10.1038/222157b0}{\emph{\nat}
  {\bfseries 222} (1969) 157}.

\bibitem{PhysRevLett.82.4964}
B.E.~Schaefer, \emph{Severe limits on variations of the speed of light with
  frequency}, \href{https://doi.org/10.1103/PhysRevLett.82.4964}{\emph{Phys.
  Rev. Lett.} {\bfseries 82} (1999) 4964}.

\bibitem{Wu_2016}
X.-F.~Wu, S.-B.~Zhang, H.~Gao, J.-J.~Wei, Y.-C.~Zou, W.-H.~Lei et~al.,
  \emph{{CONSTRAINTS} {ON} {THE} {PHOTON} {MASS} {WITH} {FAST} {RADIO}
  {BURSTS}}, \href{https://doi.org/10.3847/2041-8205/822/1/l15}{\emph{The
  Astrophysical Journal} {\bfseries 822} (2016) L15}.

\bibitem{2016PhLB..757..548B}
L.~{Bonetti}, J.~{Ellis}, N.E.~{Mavromatos}, A.S.~{Sakharov},
  E.K.~{Sarkisyan-Grinbaum} and A.D.A.M.~{Spallicci}, \emph{{Photon mass limits
  from fast radio bursts}},
  \href{https://doi.org/10.1016/j.physletb.2016.04.035}{\emph{Physics Letters
  B} {\bfseries 757} (2016) 548}
  [\href{https://arxiv.org/abs/1602.09135}{{\ttfamily 1602.09135}}].

\bibitem{2017PhLB..768..326B}
L.~{Bonetti}, J.~{Ellis}, N.E.~{Mavromatos}, A.S.~{Sakharov},
  E.K.~{Sarkisyan-Grinbaum} and A.D.A.M.~{Spallicci}, \emph{{FRB 121102 casts
  new light on the photon mass}},
  \href{https://doi.org/10.1016/j.physletb.2017.03.014}{\emph{Physics Letters
  B} {\bfseries 768} (2017) 326}
  [\href{https://arxiv.org/abs/1701.03097}{{\ttfamily 1701.03097}}].

\bibitem{ZHANG201620}
B.~Zhang, Y.-T.~Chai, Y.-C.~Zou and X.-F.~Wu, \emph{Constraining the mass of
  the photon with gamma-ray bursts},
  \href{https://doi.org/https://doi.org/10.1016/j.jheap.2016.07.001}{\emph{Journal
  of High Energy Astrophysics} {\bfseries 11-12} (2016) 20}.

\bibitem{PhysRevD.95.123010}
L.~Shao and B.~Zhang, \emph{Bayesian framework to constrain the photon mass
  with a catalog of fast radio bursts},
  \href{https://doi.org/10.1103/PhysRevD.95.123010}{\emph{Phys. Rev. D}
  {\bfseries 95} (2017) 123010}.

\bibitem{Wei_2017}
J.-J.~Wei, E.-K.~Zhang, S.-B.~Zhang and X.-F.~Wu, \emph{New limits on the
  photon mass with radio pulsars in the magellanic clouds},
  \href{https://doi.org/10.1088/1674-4527/17/2/13}{\emph{Research in Astronomy
  and Astrophysics} {\bfseries 17} (2017) 13}.

\bibitem{2018JCAP...07..045W}
J.-J.~{Wei} and X.-F.~{Wu}, \emph{{Robust limits on photon mass from
  statistical samples of extragalactic radio pulsars}},
  \href{https://doi.org/10.1088/1475-7516/2018/07/045}{\emph{\jcap} {\bfseries
  2018} (2018) 045} [\href{https://arxiv.org/abs/1803.07298}{{\ttfamily
  1803.07298}}].

\bibitem{2020RAA....20..206W}
J.-J.~{Wei} and X.-F.~{Wu}, \emph{{Combined limit on the photon mass with nine
  localized fast radio bursts}},
  \href{https://doi.org/10.1088/1674-4527/20/12/206}{\emph{Research in
  Astronomy and Astrophysics} {\bfseries 20} (2020) 206}
  [\href{https://arxiv.org/abs/2006.09680}{{\ttfamily 2006.09680}}].

\bibitem{2021FrPhy..1644300W}
J.-J.~{Wei} and X.-F.~{Wu}, \emph{{Testing fundamental physics with
  astrophysical transients}},
  \href{https://doi.org/10.1007/s11467-021-1049-x}{\emph{Frontiers of Physics}
  {\bfseries 16} (2021) 44300}
  [\href{https://arxiv.org/abs/2102.03724}{{\ttfamily 2102.03724}}].

\bibitem{Xing_2019}
N.~Xing, H.~Gao, J.-J.~Wei, Z.~Li, W.~Wang, B.~Zhang et~al., \emph{Limits on
  the weak equivalence principle and photon mass with {FRB} 121102 subpulses},
  \href{https://doi.org/10.3847/2041-8213/ab3c5f}{\emph{The Astrophysical
  Journal} {\bfseries 882} (2019) L13}.

\bibitem{2021PhLB..82036596W}
H.~{Wang}, X.~{Miao} and L.~{Shao}, \emph{{Bounding the photon mass with
  cosmological propagation of fast radio bursts}},
  \href{https://doi.org/10.1016/j.physletb.2021.136596}{\emph{Physics Letters
  B} {\bfseries 820} (2021) 136596}
  [\href{https://arxiv.org/abs/2103.15299}{{\ttfamily 2103.15299}}].

\bibitem{PhysRevLett.26.721}
E.R.~Williams, J.E.~Faller and H.A.~Hill, \emph{New experimental test of
  coulomb's law: A laboratory upper limit on the photon rest mass},
  \href{https://doi.org/10.1103/PhysRevLett.26.721}{\emph{Phys. Rev. Lett.}
  {\bfseries 26} (1971) 721}.

\bibitem{PhysRevLett.68.3383}
M.A.~Chernikov, C.J.~Gerber, H.R.~Ott and H.-J.~Gerber, \emph{Low-temperature
  upper limit of the photon mass: Experimental null test of amp\`ere's law},
  \href{https://doi.org/10.1103/PhysRevLett.68.3383}{\emph{Phys. Rev. Lett.}
  {\bfseries 68} (1992) 3383}.

\bibitem{PhysRevLett.80.1826}
R.~Lakes, \emph{Experimental limits on the photon mass and cosmic magnetic
  vector potential},
  \href{https://doi.org/10.1103/PhysRevLett.80.1826}{\emph{Phys. Rev. Lett.}
  {\bfseries 80} (1998) 1826}.

\bibitem{2003PhRvL..90h1801L}
J.~{Luo}, L.-C.~{Tu}, Z.-K.~{Hu} and E.-J.~{Luan}, \emph{{New Experimental
  Limit on the Photon Rest Mass with a Rotating Torsion Balance}},
  \href{https://doi.org/10.1103/PhysRevLett.90.081801}{\emph{\prl} {\bfseries
  90} (2003) 081801}.

\bibitem{PhysRevLett.35.1402}
L.~Davis, A.S.~Goldhaber and M.M.~Nieto, \emph{Limit on the photon mass deduced
  from pioneer-10 observations of jupiter's magnetic field},
  \href{https://doi.org/10.1103/PhysRevLett.35.1402}{\emph{Phys. Rev. Lett.}
  {\bfseries 35} (1975) 1402}.

\bibitem{1997PPCF...39...73R}
D.D.~{Ryutov}, \emph{{The role of finite photon mass in magnetohydrodynamics of
  space plasmas}},
  \href{https://doi.org/10.1088/0741-3335/39/5A/008}{\emph{Plasma Physics and
  Controlled Fusion} {\bfseries 39} (1997) A73}.

\bibitem{2007PPCF...49..429R}
D.D.~{Ryutov}, \emph{{Using plasma physics to weigh the photon}},
  \href{https://doi.org/10.1088/0741-3335/49/12B/S40}{\emph{Plasma Physics and
  Controlled Fusion} {\bfseries 49} (2007) B429}.

\bibitem{2016APh....82...49R}
A.~{Retin{\`o}}, A.D.A.M.~{Spallicci} and A.~{Vaivads}, \emph{{Solar wind test
  of the de Broglie-Proca massive photon with Cluster multi-spacecraft data}},
  \href{https://doi.org/10.1016/j.astropartphys.2016.05.006}{\emph{Astroparticle
  Physics} {\bfseries 82} (2016) 49}
  [\href{https://arxiv.org/abs/1302.6168}{{\ttfamily 1302.6168}}].

\bibitem{1959PThPS..11....1Y}
Y.~{Yamaguchi}, \emph{{A Composite Theory of Elementary Particles}},
  \href{https://doi.org/10.1143/PTPS.11.1}{\emph{Progress of Theoretical
  Physics Supplement} {\bfseries 11} (1959) 1}.

\bibitem{1976UsFiN.119..551C}
G.V.~{Chibisov}, \emph{{Astrophysical upper limits for the rest mass of a
  photon.}}, {\emph{Uspekhi Fizicheskikh Nauk} {\bfseries 119} (1976) 551}.

\bibitem{2007PhRvL..98a0402A}
E.~{Adelberger}, G.~{Dvali} and A.~{Gruzinov}, \emph{{Photon-Mass Bound
  Destroyed by Vortices}},
  \href{https://doi.org/10.1103/PhysRevLett.98.010402}{\emph{\prl} {\bfseries
  98} (2007) 010402} [\href{https://arxiv.org/abs/hep-ph/0306245}{{\ttfamily
  hep-ph/0306245}}].

\bibitem{PhysRevD.8.2349}
D.D.~Lowenthal, \emph{Limits on the photon mass},
  \href{https://doi.org/10.1103/PhysRevD.8.2349}{\emph{Phys. Rev. D} {\bfseries
  8} (1973) 2349}.

\bibitem{PhysRevD.69.107501}
A.~Accioly and R.~Paszko, \emph{Photon mass and gravitational deflection},
  \href{https://doi.org/10.1103/PhysRevD.69.107501}{\emph{Phys. Rev. D}
  {\bfseries 69} (2004) 107501}.

\bibitem{PhysRevLett.109.131102}
P.~Pani, V.~Cardoso, L.~Gualtieri, E.~Berti and A.~Ishibashi, \emph{Black-hole
  bombs and photon-mass bounds},
  \href{https://doi.org/10.1103/PhysRevLett.109.131102}{\emph{Phys. Rev. Lett.}
  {\bfseries 109} (2012) 131102}.

\bibitem{Yang_2017}
Y.-P.~Yang and B.~Zhang, \emph{Tight constraint on photon mass from pulsar
  spindown}, \href{https://doi.org/10.3847/1538-4357/aa74de}{\emph{The
  Astrophysical Journal} {\bfseries 842} (2017) 23}.

\bibitem{Hankins_2007}
T.H.~Hankins and J.A.~Eilek, \emph{Radio emission signatures in the crab
  pulsar}, \href{https://doi.org/10.1086/522362}{\emph{The Astrophysical
  Journal} {\bfseries 670} (2007) 693}.

\bibitem{2021ApJ...911L...3P}
Z.~{Pleunis}, D.~{Michilli}, C.G.~{Bassa}, J.W.T.~{Hessels}, A.~{Naidu},
  B.C.~{Andersen} et~al., \emph{{LOFAR Detection of 110-188 MHz Emission and
  Frequency-dependent Activity from FRB 20180916B}},
  \href{https://doi.org/10.3847/2041-8213/abec72}{\emph{\apjl} {\bfseries 911}
  (2021) L3} [\href{https://arxiv.org/abs/2012.08372}{{\ttfamily 2012.08372}}].

\bibitem{2020A&A...641A...6P}
{Planck Collaboration}, N.~{Aghanim}, Y.~{Akrami}, M.~{Ashdown}, J.~{Aumont},
  C.~{Baccigalupi} et~al., \emph{{Planck 2018 results. VI. Cosmological
  parameters}}, \href{https://doi.org/10.1051/0004-6361/201833910}{\emph{\aap}
  {\bfseries 641} (2020) A6}
  [\href{https://arxiv.org/abs/1807.06209}{{\ttfamily 1807.06209}}].

\bibitem{1979rpa..book.....R}
G.B.~{Rybicki} and A.P.~{Lightman}, \emph{{Radiative processes in
  astrophysics}} (1979).

\bibitem{Deng_2014}
W.~Deng and B.~Zhang, \emph{{COSMOLOGICAL} {IMPLICATIONS} {OF} {FAST} {RADIO}
  {BURST}/{GAMMA}-{RAY} {BURST} {ASSOCIATIONS}},
  \href{https://doi.org/10.1088/2041-8205/783/2/l35}{\emph{The Astrophysical
  Journal} {\bfseries 783} (2014) L35}.

\bibitem{2004ApJ...612..375C}
J.M.~{Cordes}, N.D.R.~{Bhat}, T.H.~{Hankins}, M.A.~{McLaughlin} and J.~{Kern},
  \emph{{The Brightest Pulses in the Universe: Multifrequency Observations of
  the Crab Pulsar's Giant Pulses}},
  \href{https://doi.org/10.1086/422495}{\emph{\apj} {\bfseries 612} (2004) 375}
  [\href{https://arxiv.org/abs/astro-ph/0304495}{{\ttfamily
  astro-ph/0304495}}].

\bibitem{2005AdSpR..35.1166J}
A.~{Jessner}, A.~{S{\l}owikowska}, B.~{Klein}, H.~{Lesch}, C.H.~{Jaroschek},
  G.~{Kanbach} et~al., \emph{{Giant radio pulses from the Crab pulsar}},
  \href{https://doi.org/10.1016/j.asr.2005.01.091}{\emph{Advances in Space
  Research} {\bfseries 35} (2005) 1166}
  [\href{https://arxiv.org/abs/astro-ph/0410003}{{\ttfamily
  astro-ph/0410003}}].

\bibitem{2010A&A...524A..60J}
A.~{Jessner}, M.V.~{Popov}, V.I.~{Kondratiev}, Y.Y.~{Kovalev}, D.~{Graham},
  A.~{Zensus} et~al., \emph{{Giant pulses with nanosecond time resolution
  detected from the Crab pulsar at 8.5 and 15.1 GHz}},
  \href{https://doi.org/10.1051/0004-6361/201014806}{\emph{\aap} {\bfseries
  524} (2010) A60} [\href{https://arxiv.org/abs/1008.3992}{{\ttfamily
  1008.3992}}].

\bibitem{2001ApJ...557L..93R}
R.W.~{Romani} and S.~{Johnston}, \emph{{Giant Pulses from the Millisecond
  Pulsar B1821-24}}, \href{https://doi.org/10.1086/323415}{\emph{\apjl}
  {\bfseries 557} (2001) L93}
  [\href{https://arxiv.org/abs/astro-ph/0107511}{{\ttfamily
  astro-ph/0107511}}].

\bibitem{2003ApJ...590L..95J}
S.~{Johnston} and R.W.~{Romani}, \emph{{Giant Pulses from PSR B0540-69 in the
  Large Magellanic Cloud}}, \href{https://doi.org/10.1086/376826}{\emph{\apjl}
  {\bfseries 590} (2003) L95}
  [\href{https://arxiv.org/abs/astro-ph/0305235}{{\ttfamily
  astro-ph/0305235}}].

\bibitem{2004ApJ...616..439S}
V.A.~{Soglasnov}, M.V.~{Popov}, N.~{Bartel}, W.~{Cannon}, A.Y.~{Novikov},
  V.I.~{Kondratiev} et~al., \emph{{Giant Pulses from PSR B1937+21 with Widths
  $<=15$ Nanoseconds and T$_{b}$$>=5\times$ 10$^{39}$ K, the Highest
  Brightness Temperature Observed in the Universe}},
  \href{https://doi.org/10.1086/424908}{\emph{\apj} {\bfseries 616} (2004) 439}
  [\href{https://arxiv.org/abs/astro-ph/0408285}{{\ttfamily
  astro-ph/0408285}}].

\bibitem{2005ApJ...625..951K}
H.S.~{Knight}, M.~{Bailes}, R.N.~{Manchester} and S.M.~{Ord}, \emph{{A Search
  for Giant Pulses from Millisecond Pulsars}},
  \href{https://doi.org/10.1086/429533}{\emph{\apj} {\bfseries 625} (2005) 951}
  [\href{https://arxiv.org/abs/astro-ph/0502203}{{\ttfamily
  astro-ph/0502203}}].

\bibitem{2020Natur.577..190M}
B.~{Marcote}, K.~{Nimmo}, J.W.T.~{Hessels}, S.P.~{Tendulkar}, C.G.~{Bassa},
  Z.~{Paragi} et~al., \emph{{A repeating fast radio burst source localized to a
  nearby spiral galaxy}},
  \href{https://doi.org/10.1038/s41586-019-1866-z}{\emph{\nat} {\bfseries 577}
  (2020) 190} [\href{https://arxiv.org/abs/2001.02222}{{\ttfamily
  2001.02222}}].

\end{thebibliography}

\end{document}